\documentclass[ notitlepage,amsmath,amssymb,preprintnumbers,superscriptaddress,nofootinbib,aps,twocolumn,prd,floatfix]{revtex4-1}

\usepackage{amssymb,amsmath,bm,tensor,braket}
\usepackage[colorlinks]{hyperref}
\usepackage[T1]{fontenc} 
\usepackage{graphicx}
\usepackage{epstopdf}
\usepackage{makecell}
\graphicspath{{figures/}}
\usepackage{natbib}
\usepackage{graphicx}
\usepackage{dcolumn}
\usepackage{bm}
\usepackage{xcolor}
\usepackage{physics}    
\usepackage{siunitx}    

\usepackage{soul}

\begin{document}

\preprint{CPTNP-2025-022}

\title{Probing top-quark electroweak couplings indirectly at the Electron-Ion Collider}

\author{Xu-Hui Jiang} 
\thanks{These authors contributed equally to this work.}
\affiliation{Institute of High Energy Physics, Chinese Academy of Sciences, Beijing 100049, China}
\affiliation{China Center of Advanced Science and Technology, Beijing 100190, China}

\author{Yiming Liu} 
\thanks{These authors contributed equally to this work.}
\affiliation{School of Physics, Beijing Institute of Technology, Beijing 100081, China}

\author{Bin Yan} \email{yanbin@ihep.ac.cn}
\affiliation{Institute of High Energy Physics, Chinese Academy of Sciences, Beijing 100049, China}
\affiliation{Center for High Energy Physics, Peking University, Beijing 100871, China}


\begin{abstract}
Top quark electroweak interactions serve as a sensitive probe for beyond-Standard-Model physics, potentially exhibiting significant deviations from Standard Model predictions and offering crucial tests of fundamental symmetries in ultraviolet physics. This Letter investigates their measurement through deep inelastic scattering (DIS) processes involving top quark loops at the Electron-Ion Collider (EIC), within the Standard Model Effective Field Theory framework. We demonstrate that left-handed electron beam polarization can significantly enhance DIS cross sections for these interactions compared to right-handed scenario, highlighting the pivotal role of high polarization at the EIC in constraining these couplings. Compared to direct $pp \to t\bar{t}Z$ measurements at the Large Hadron Collider (LHC), DIS measurements could significantly improve these coupling constraints and effectively resolve parameter space degeneracies present in LHC data. Moreover, the expected precision from the EIC's high-luminosity phase matches that of LEP electroweak precision measurements, underscoring the EIC's significant potential to probe top quark properties.
\end{abstract}

\maketitle

\emph{Introduction.---}
Top quark, regarded as the heaviest elementary particle in the Standard Model (SM), has a mass near the electroweak scale. It also acts as the only fermion with an $\mathcal{O}(1)$ Yukawa coupling. Consequently, it provides a unique window into both SM dynamics and potential new physics beyond the SM (BSM). Its special properties make it crucial for studying electroweak symmetry breaking, vacuum stability, and the hierarchy problem. Additionally, it also serves as a sensitive probe for BSM scenarios like supersymmetry and composite Higgs models that often predict top partners. 

However, the absence of new heavy resonance signals at the Large Hadron Collider (LHC) also underscores the importance of complementary approaches to probe potential BSM effects. One promising direction is precision measurements of SM processes, including total cross sections and differential distributions, which may reveal subtle deviations indicative of BSM dynamics. A powerful model-independent framework to characterize such deviations is the SM Effective Field Theory  (SMEFT). It extends the SM by incorporating higher-dimensional operators, constructed from SM fields with gauge symmetry $SU(3)_C\otimes SU(2)_L\otimes U(1)_Y$, and  suppressed by powers of a high-energy scale $\Lambda$. These operators encode the effects of heavy, integrated-out degrees of freedom into Wilson coefficients (WCs), providing a systematic way to parameterize BSM contributions at the TeV scale. Up to dimensional-six level,  the SMEFT Lagrangian takes the following form:
\begin{equation}
    \mathcal L = \mathcal L_{\rm SM} + \sum_i \frac{C_i}{\Lambda^2} \mathcal O_i~,
\end{equation}
where $ \mathcal L_{\rm SM} $ represents the SM Lagrangian, $\mathcal O_i$ are the dimension-six operators, and  $C_i$ denotes their associated WCs~\cite{Buchmuller:1985jz,Grzadkowski:2010es}.  

The LHC has demonstrated remarkable capability as a precision top-quark factory, enabling detailed measurements of top-quark production cross sections across various channels and providing stringent constraints on top-quark couplings~\cite{Bernreuther:2015yna, CMS:2020lrr,ATLAS:2024hmk,CMS:2022lmh,ATLAS:2023eld,CMS:2019too,CMS:2021aly, Bernreuther:2024ltu}. With the upcoming high-luminosity upgrade (HL-LHC), which will deliver an integrated luminosity of up to $3000~\text{fb}^{-1}$ , the event yields are expected to increase by approximately a factor of 20 compared to LHC Run II. This significant enhancement in statistics will substantially improve the sensitivity to potential BSM effects.  Additionally, lepton colliders can provide complementary constraints on top-quark properties through precision electroweak measurements, as demonstrated at LEP, where top-quark effects manifest via quantum loops. Future high-energy lepton colliders, such as CEPC~\cite{CEPCStudyGroup:2018ghi, CEPCStudyGroup:2023quu}, FCC-ee~\cite{FCC:2018evy}, CLIC~\cite{Linssen:2012hp}, and ILC~\cite{ILC:2007bjz, ILC:2013jhg}, also enable promising studies of top-quark physics. The potential of these facilities for precision top-quark measurements has been thoroughly explored in recent studies~\cite{Cao:2015qta,Durieux:2018tev, Liu:2022vgo, Celada:2024mcf, Cornet-Gomez:2025jot, deBlas:2022ofj, Bissmann:2020mfi, Durieux:2019rbz, Ellis:2020unq, Garosi:2023yxg, Altmann:2025feg}. A global approach, combining direct collider measurements with precision electroweak data, provides particularly robust constraints on BSM effects~\cite{Efrati:2015eaa,Cao:2015doa,Schulze:2016qas,Durieux:2019rbz,Cao:2020npb,Ellis:2020unq, Liu:2022vgo, Bissmann:2020mfi,Cao:2021wcc, Garosi:2023yxg,Altmann:2025feg}.

\begin{figure}
    \centering
    \includegraphics[width=\linewidth]{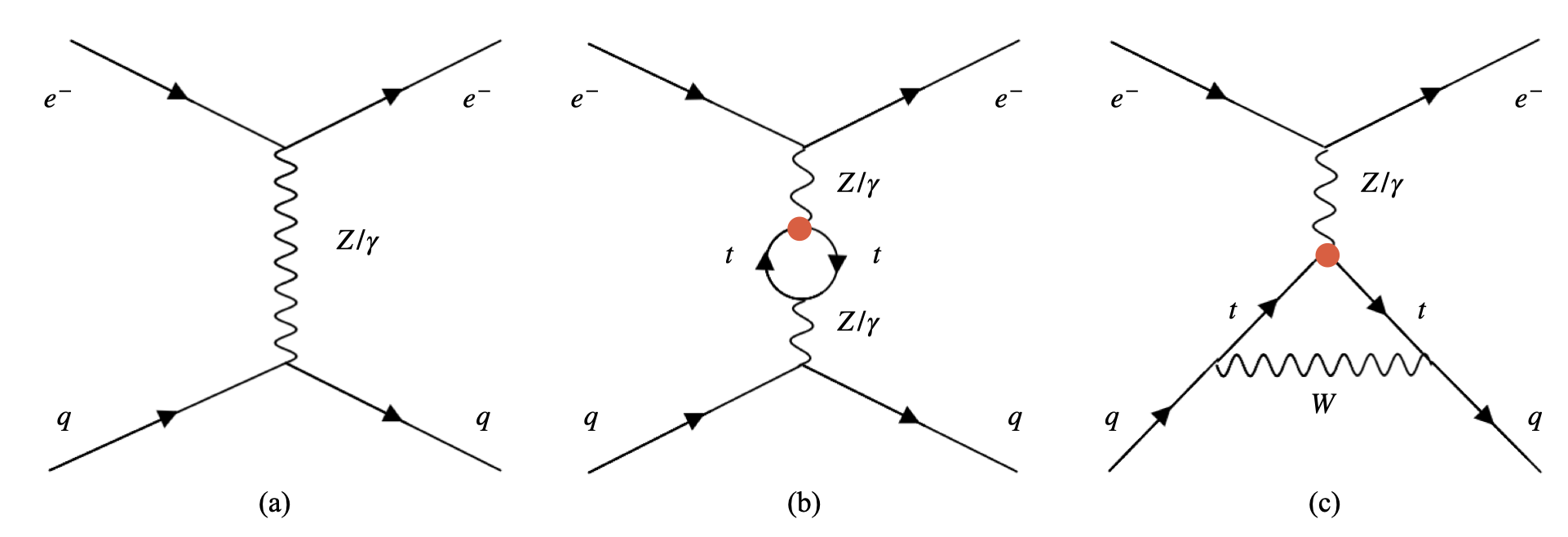}
    \caption{Typical Feynman diagrams for $e^- p \to e^- j$ at the parton level, depicting (a) tree-level contribution, (b) NLO contribution with a self-energy correction, and (c) NLO contribution with a penguin diagram. The red dot indicates deviations of top-quark couplings with gauge bosons from SM predictions.}
    \label{fig:efef}
\end{figure}

In this Letter, we investigate the potential of the upcoming Electron-Ion Collider (EIC)~\cite{AbdulKhalek:2021gbh}  to probe top-quark electroweak couplings  through quantum loop effects at next-to-leading order (NLO) via precision measurements of deep inelastic scattering (DIS) $e^- p \to e^- j$ (see Fig.~\ref{fig:efef}) at $\sqrt{s}=100~{\rm GeV}$ ($E_e = 10~\text{GeV}$ and $E_p = 270~\text{GeV}$), considering both baseline ($\mathcal{L} = 100~\text{fb}^{-1}$) and high-luminosity ($\mathcal{L} = 1000~\text{fb}^{-1}$) scenarios~\cite{AbdulKhalek:2021gbh} . While the EIC is primarily designed for nuclear structure studies~\cite{AbdulKhalek:2021gbh}, recent theoretical works~\cite{Gonderinger:2010yn,Boughezal:2020uwq,Cirigliano:2021img,Yan:2021htf,Liu:2021lan,Li:2021uww,Davoudiasl:2021mjy,Yan:2022npz,Zhang:2022zuz,Batell:2022ogj,Boughezal:2022pmb,AbdulKhalek:2022hcn,Davoudiasl:2023pkq,Boughezal:2023ooo,Delzanno:2024ooj,Wang:2024zns,Wen:2024cfu,Du:2024sjt,Gao:2024rgl,Balkin:2023gya, Deng:2025hio,Davoudiasl:2025rpn,Bellafronte:2025ubi}  highlight its significant capabilities for precision SM tests and BSM physics searches.  By leveraging the high polarization of the EIC's electron beam, we demonstrate that these DIS measurements can significantly improve constraints on top-quark gauge couplings and help resolve degeneracies in the parameter space of WCs~\footnote{The polarization of the proton beam is not expected to significantly affect our predictions, as polarized PDFs are strongly suppressed relative to their unpolarized counterparts~\cite{Ball:2013lla}.}. 

\emph{Top quark operators in SMEFT.---}
Following the parameterization of Ref.~\cite{Aguilar-Saavedra:2018ksv}, we consider the following dimension-six operators involving top quark gauge couplings in our analysis: 
\begin{align}
\mathcal O_{Ht} & = i(H^\dagger \overleftrightarrow D_\mu H) (\bar t_R \gamma^\mu t_R)~, \\
\mathcal O_{Hq}^{(1)} & = i(H^\dagger \overleftrightarrow D_\mu H) (\bar q_L \gamma^\mu  q_L)~, \\
\mathcal O_{Hq}^{(3)} & = i(H^\dagger \overleftrightarrow D_\mu^I H) (\bar q_L \gamma^\mu \sigma^I q_L)~,\\
\mathcal O_{tW} & = (\bar q_L \sigma^{\mu\nu} \sigma^I t_R) \tilde H W_{\mu\nu}^I + h.c.~, \\
\mathcal O_{tB} & = (\bar q_L \sigma^{\mu\nu}  t_R) \tilde H B_{\mu\nu} + h.c. ~,
\end{align}
where $q_L=(t_L,b_L)^T$ represents the $SU(2)_L$ weak doublet of the third generation left-handed quark fields. $t_R$ is $SU(2)_L$ weak singlet of right-handed top quark field. We also have $H^\dagger \overleftrightarrow D_\mu H\equiv  H^\dagger D_\mu H-(D_\mu H)^\dagger H$, with $H$ being the Higgs field, $D_\mu$ being the covariant derivative, and $\tilde H=i\sigma^2H^*$ being the Higgs field in the conjugate representation.
$W_{\mu\nu}^I$ and $B_{\mu\nu}$ denote the field strength of electroweak gauge bosons, and $\sigma^I$ are the Pauli matrices. 
For our analysis, we define the linear combination of operators $\mathcal{O}_{HQ}^{(\pm)} = \mathcal{O}_{Hq}^{(1)} \pm \mathcal{O}_{Hq}^{(3)}$. The operator $\mathcal{O}_{HQ}^{(+)}$ contributes to the $b_L \bar{b}_L Z$ coupling, which is tightly constrained by LEP data~\cite{Choudhury:2001hs,ALEPH:2005ab,Yan:2021veo,Yan:2021htf,Li:2021uww,Dong:2022ayy,Yan:2023ccj} and is therefore excluded from this study. 

After the electroweak symmetry breaking $\langle H\rangle=v/\sqrt{2}$ with $v=246~{\rm GeV}$, these operators generate the following effective top quark couplings to gauge bosons:
\begin{align}
 \mathcal L_{\rm eff}&=\frac{g}{2c_W}\bar{t}\left[\gamma^\mu \left(g_L P_L+g_R P_R\right)\right]tZ_\mu\nonumber\\
 &+ \bar{t}\left[\sigma^{\mu\nu} \left(f_{Z}^V +i f_{Z}^A \gamma^5\right)\right]tZ_{\mu\nu} \nonumber \\
 &+ \bar{t}\left[\sigma^{\mu\nu} \left(f_{A}^V +i f_{A}^A \gamma^5\right)\right]tA_{\mu\nu}~,
\end{align}
where $Z_{\mu\nu}$ and $A_{\mu\nu}$ are the field strength tensors of the $Z$ boson and photon, respectively, and  the coefficients can be expressed as the WCs of top quark operators,
\begin{align}
g_L &= -2\frac{v^2}{\Lambda^2}C_{HQ}^{(-)}=-\frac{v^2}{\Lambda^2}(C_{Hq}^{(1)}-C_{Hq}^{(3)})~, \\
g_R & =-\frac{v^2}{\Lambda^2}C_{Ht}~,\\
f_{Z(A)}^{V} & = \frac{v}{\sqrt 2 \Lambda^2} C_{tZ(A)}~,\\
f_{Z(A)}^{A} & = \frac{v}{\sqrt 2 \Lambda^2} C_{tZ(A)}^A~,\\
C_{tZ} &= \text{Re}(c_W C_{tW}-s_W C_{tB})~,\\
C_{tZ}^A &= \text{Im}(c_W C_{tW}-s_W C_{tB})~,\\
C_{tA} &= \text{Re}(s_W C_{tW}+c_W C_{tB})~,\\
C_{tA}^A &= \text{Im}(s_W C_{tW}+c_W C_{tB})~.
\end{align}
Here we have $c_W\equiv \cos\theta_W$ and $s_W\equiv \sin\theta_W$, with $\theta_W$ being the weak mixing angle of the SM.
The imaginary components of the Wilson coefficients $C_{tZ}^A$ and $C_{tA}^A$, which are CP-violating terms, result in vanishing contributions to the inclusive cross section at $\mathcal{O}(1/\Lambda^2)$ due to cancellations stemming from their CP-odd properties. 

As shown in Fig.~\ref{fig:efef}, the anomalous $t\bar{t} Z$ and $t\bar{t}\gamma$ couplings primarily contribute to the DIS process through self-energy corrections (Fig.~\ref{fig:efef} (b)). Penguin diagram contributions (Fig.~\ref{fig:efef} (c)) are negligible due to suppression by the bottom quark parton distribution function (PDF) or off-diagonal CKM matrix elements. Consequently, we adopt a four-flavor scheme in our calculations and safely neglect penguin diagram contributions in this work. 
However, self-energy corrections from the top quark are not the sole contributors to the DIS cross sections. As demonstrated in Ref.~\cite{Zhang:2012cd}, new operators introduce shifts in renormalized parameters, such as the precisely measured input parameters $\alpha$, $m_Z$, and $G_F$. Specifically, the top operators $\mathcal{O}_{Hq}^{(3)}$ and $\mathcal{O}_{tW}$ modify the $W$ boson mass, thereby affecting the Fermi constant through renormalization. Although $W$-loop contributions are not directly accessible in the DIS process of Fig.~\ref{fig:efef}, these operators induce sizeable additional corrections. Notably, these effects are not captured by a simple matching between four-fermion operators in the Low-Energy Effective Field Theory (LEFT) and the top quark operators in SMEFT, requiring separate consideration in the LEFT framework~\cite{Dekens:2019ept}.

The contributions of top quark operators to the inclusive total cross section can be expressed as,
\begin{equation}\label{eq:xsec}
    \sigma = \sigma_{\text{SM}} + \sum_i r_i C_i + \sum_{i,j} r_{ij} C_i C_j~,
\end{equation}
where $C_i$ represents the WCs of the corresponding operators, and $r_i$ and $r_{ij}$ are the coefficients of the linear and quadratic terms, respectively. In Eq.~\eqref{eq:xsec}, the second term arises from interference between the SM and BSM operators, while the third term reflects correlations among these operators. The coefficients $r_i$ and $r_{ij}$ are expected to scale as $\mathcal{O}(1/\Lambda^2)$ and $\mathcal{O}(1/\Lambda^4)$, respectively. Consequently, the linear contribution typically dominates, unless a specific mechanism significantly suppresses the interference term. For the DIS process, we therefore neglect the $\mathcal{O}(1/\Lambda^4)$ contributions. Since the $r_i$ coefficients arise from NLO corrections, they depend on the SMEFT renormalization scale $\mu_{\text{EFT}}$ in the $\overline{\text{MS}}$ scheme. As noted in Ref.~\cite{Zhang:2012cd}, direct loop calculations reveal that operators contribute a finite term at NLO, which is not fully captured by renormalization group equations (RGE)~\cite{Hartmann:2015oia, Maltoni:2016yxb, Vryonidou:2018eyv, Asteriadis:2024xuk}. The coefficient $r_i$ can be parameterized as
\begin{equation}\label{eq:ri}
    r_i = \sigma_{i0} + \sigma_{i1} \log \frac{\mu_{\text{EFT}}^2}{Q^2} + \mathcal{O}\left( \log^2 \frac{\mu_{\text{EFT}}^2}{Q^2} \right)~,
\end{equation}
where $Q$ is the typical energy scale of the process, and $\sigma_{i0}$ and $\sigma_{i1}$ are coefficients independent of the renormalization scale. Here, $\sigma_{i0}$ originates from direct loop computations and is absent when only RGE are considered, while $\sigma_{i1}$ corresponds to the leading logarithmic term consistent with RGE.

\emph{DIS process at the EIC.---}
We have shown that the NLO electroweak corrections from top quark operators to the DIS process are dominated by oblique corrections. Consequently, the cross section can be conveniently expressed in the "star scheme" using the fine structure constant  $\alpha$, $Z$ boson mass $m_Z$, and $s_W$~\cite{Peskin:1991sw,Zhang:2012cd}. This requires the following substitutions, which incorporate both the direct contributions from the top quark loop, as shown in Fig.~\ref{fig:efef}(b), and the indirect corrections due to shifts in the renormalized parameters~\cite{Zhang:2012cd},
\begin{widetext}
\begin{eqnarray}%
\alpha&\rightarrow&\alpha_*=\alpha+\delta\alpha
=\alpha\left(1-\Pi'_{\gamma\gamma}(q^2)+\Pi'_{\gamma\gamma}(0)\right)
\times\left[
1-\frac{d}{dq^2}\Pi_{ZZ}(q^2)|_{q^2=m_Z^2}+\Pi'_{\gamma\gamma}(q^2)+\frac{c_W^2-s_W^2}{s_Wc_W}\Pi'_{\gamma Z}(q^2)
\right]~,
\label{replace3}\\
m_Z^2&\rightarrow&m_{Z*}^2=m_Z^2+\delta m_Z^2
=m_Z^2+\Pi_{ZZ}(m_Z^2)-\Pi_{ZZ}(q^2)+(q^2-m_Z^2)\frac{d}{dq^2}\Pi_{ZZ}(q^2)|_{q^2=m_Z^2}~,\label{replace35}
\\
s_{W}^2&\rightarrow&s_{W*}^2=s_{W}^2+\delta s_{W}^2
=s_W^2\left[
1+\frac{c_W}{s_W}\Pi'_{\gamma Z}(q^2)+\frac{c_W^2}{c_W^2-s_W^2}\left(
\Pi'_{\gamma\gamma}(0)+\frac{1}{m_W^2}\Pi_{WW}(0)-\frac{1}{m_Z^2}\Pi_{ZZ}(m_Z^2)
\right)
\right]~,\label{replace4}
\end{eqnarray}%
\end{widetext}
where $\Pi_{VV}(q^2)$ denotes the self-energy correction for gauge bosons $V = \gamma, W, Z$, and $\Pi'_{VV}(q^2) \equiv \left[ \Pi_{VV}(q^2) - \Pi_{VV}(0) \right] / q^2$. The explicit expressions for these self-energy corrections due to top quark operators are provided in Ref.~\cite{Liu:2022vgo}. The terms $\Pi'_{VV}(0)$, $\Pi_{VV}(m_Z^2)$, and $\Pi'_{VV}(m_Z^2)$ represent indirect corrections arising from shifts in the renormalized parameters. These indirect effects lead to partial cancellations with direct contributions in the parameters $\delta\alpha$ and $\delta m_Z^2$, with the corrections to $\alpha$ being further suppressed by a factor of $\alpha$ itself. Consequently, the corrections to $\delta s_W^2$ dominate the final electroweak corrections to the DIS cross-section. 
Specifically, we find that the corrections to $\delta s_W^2$ are primarily driven by indirect contributions for the operators $\mathcal{O}_{HQ}^{(-)}$, $\mathcal{O}_{Ht}$, and $\mathcal{O}_{tB}$, while direct contributions from the top quark loop dominate for the operator $\mathcal{O}_{tW}$. To validate our analytical results, we performed numerical calculations and compared them with simulations using MadGraph5\_aMC@NLO~\cite{Alwall:2014hca}, finding an excellent agreement.

\emph{Numerical results and discussion.---}
Next, we perform a detailed Monte Carlo simulation with a reweighting method~\cite{Mattelaer:2016gcx} to investigate the sensitivity of the process $e^- p \to e^- j$ at the EIC to top quark operators. The simulation adopts benchmark collider energies of $E_e = 10~\text{GeV}$ and $E_p = 270~\text{GeV}$, utilizing the "NNPDF2.3 LO" PDF~\cite{Ball:2012cx}. The following kinematic requirements are applied to the final-state particles: $p_{Tj} > 20~\text{GeV}$, $|\eta_j| < 5$, $p_{Te} > 10~\text{GeV}$, and, $|\eta_e| < 2.5$.
Although the EIC operates at significantly lower energies than the LHC, its high lepton beam polarization offers a unique opportunity to enhance the precision of top quark electroweak coupling measurements. To explore this potential, we consider three electron beam polarization scenarios: $P_e = +70\%$, $P_e = -70\%$, and $P_e = 0$ (unpolarized). The following input parameters are used in our analysis: $\alpha^{-1}(m_{Z})=127.9$, $m_{Z}=91.1876~\text{GeV}$, $m_{t}=172.5~\text{GeV}$, $G_F=1.166379\times 10^{-5}~\text{GeV}^{-2}$ and $\Lambda=1~{\rm TeV}$. Table~\ref{tab: xsec1} presents the coefficients $r_i$ in the cross-section expression, as given in Eq.~\eqref{eq:xsec}, obtained from our calculation. The $t \bar{t} \gamma$ anomalous couplings have been tightly constrained by the ATLAS~\cite{ATLAS:2024hmk} and CMS~\cite{CMS:2022lmh} collaborations through the $pp \to t \bar{t} \gamma$ process at the LHC, with constraints approximately one order of magnitude stronger than those on $t \bar{t} Z$ couplings~\cite{ATLAS:2023eld,CMS:2019too,CMS:2021aly}. Consequently, we exclude their contributions in this analysis, despite the coefficient $r_{t\gamma}$ being several times larger than $r_{tZ}$ at the EIC. To enable comparison with constraints from LEP data~\cite{Liu:2022vgo}, we set the EFT renormalization scale to $\mu_{\text{EFT}} = 125~{\rm GeV}$. Notably, the cross sections for right-handed polarized electron beam are suppressed by approximately one order of magnitude compared to those for left-handed polarized beam, as shown in Table~\ref{tab: xsec1}. This difference arises from both the chiral structure of the $Ze\bar{e}$ coupling and the indirect effects of gauge-boson self-energies, which effectively shift the weak mixing angle. This significant enhancement emphasizes the critical role of a left-handed polarized electron beam in probing top quark gauge couplings.

\begin{table}[htbp]
\centering
\begin{tabular}{cccccc}
\hline
 $P_e$ & $\sigma_{\rm SM}$[pb] & $r_{Ht}$[pb] & $r_{HQ}^{(-)}$[pb] & $r_{tZ}$[pb]  \\
\hline
$-70\%$ & $96.3$ & $3.39\times 10^{-2}$ & $-3.06\times 10^{-2}$ & $1.47\times 10^{-2}$  \\
\hline
$0$ & $92.9$ & $1.62\times 10^{-2}$ & $-1.46\times 10^{-2}$ & $6.18\times 10^{-3}$  \\
\hline
$+70\%$ & $89.7$ & $-1.76\times 10^{-3}$ & $1.65\times 10^{-3}$ & $-2.40\times 10^{-3}$ \\
\hline
\end{tabular}
\caption{Coefficients $r_i$ in the cross-section expression (Eq.~\eqref{eq:xsec}) for the DIS process at the EIC with $\sqrt{s} = \SI{100}{GeV}$, evaluated for three electron beam polarization scenarios: $P_e = \pm 70\%$ and $P_e = 0$.}\label{tab: xsec1}
\end{table}

To determine the WCs of top quark operators, we perform a $\chi^2$ analysis,
\begin{equation}
    \chi^2(P_e) =\frac{(\sum_i r_i C_i)^2}{\delta_{\rm stat}^2}= \frac{(\sum_i r_i C_i)^2 \mathcal L}{\sigma_{\rm SM}^{(P_e)}}~,
\end{equation}
where the experimental values of the cross sections are assumed to be the same as the SM predictions, and a Gaussian posterior distribution is adopted~\footnote{Additional kinematic observables may offer improved sensitivity, but a detailed study of these is left for future work.}. Systematic uncertainties are expected to be well controlled at future EIC experiments due to sophisticated studies of inclusive backgrounds. We consider integrated luminosities of $\mathcal{L}=100~{\rm fb}^{-1}$  and $\mathcal{L}=1000~{\rm fb}^{-1}$  (HL-phase) in the following numerical analysis.

\begin{figure*}
    \centering
    \includegraphics[width=0.3\linewidth]{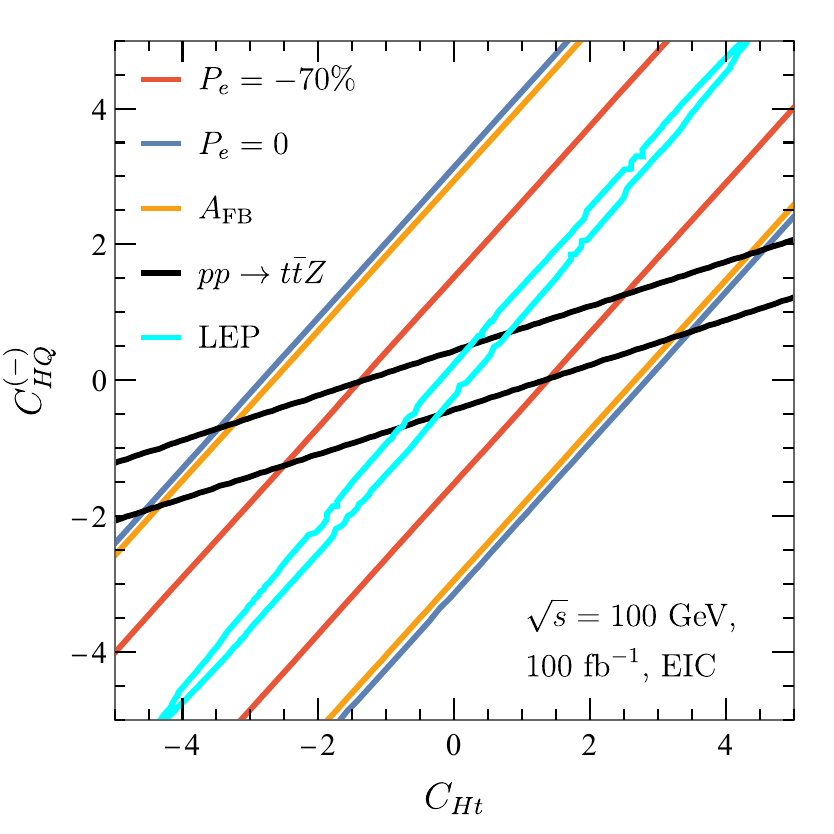}
    \includegraphics[width=0.3\linewidth]{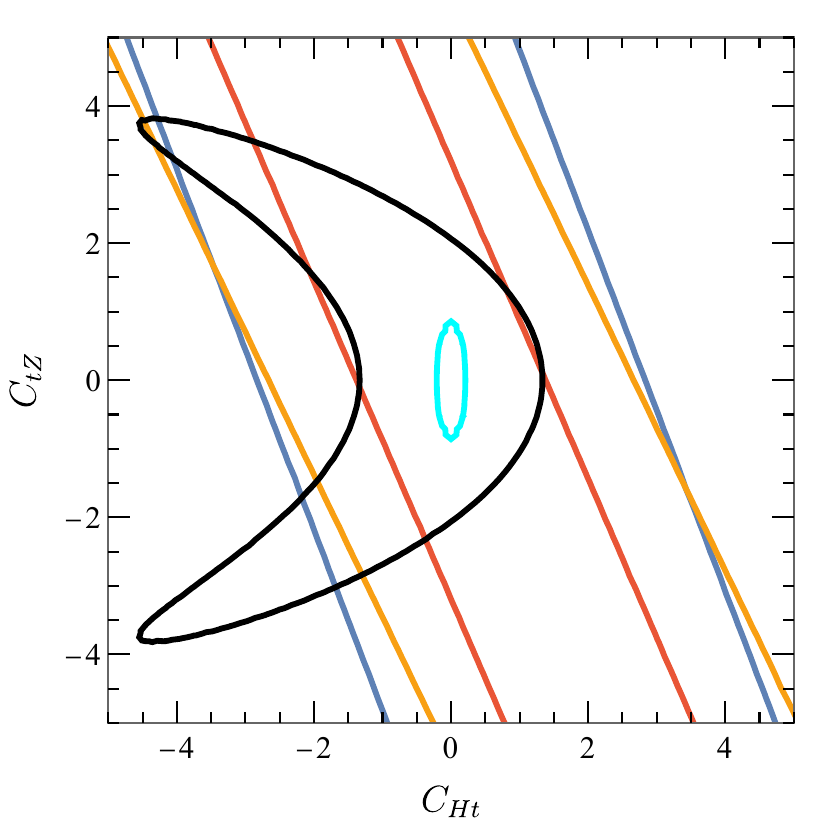}
    \includegraphics[width=0.3\linewidth]{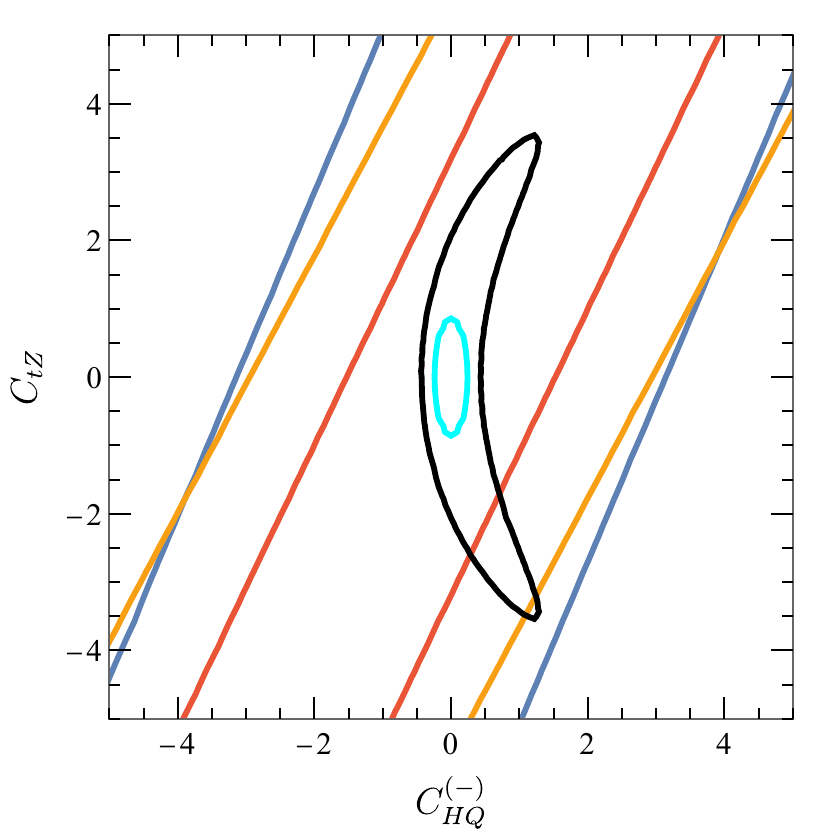}
    \caption{Expected constraints on top quark electroweak couplings at 68\% confidence level from the EIC at $\sqrt{s} = 100~\rm GeV$ with $P_e = -70\%$ (red lines), $P_e = 0$ (blue lines), and $A_{\rm FB}$ (orange lines), assuming an integrated luminosity of $100~\rm fb^{-1}$. For comparison, bounds from the LHC~\cite{ATLAS:2023eld} and LEP~\cite{Liu:2022vgo} are shown in black and cyan, respectively.}
    \label{fig: lhc_eic100}
\end{figure*}

\begin{figure*}
    \centering
    \includegraphics[width=0.3\linewidth]{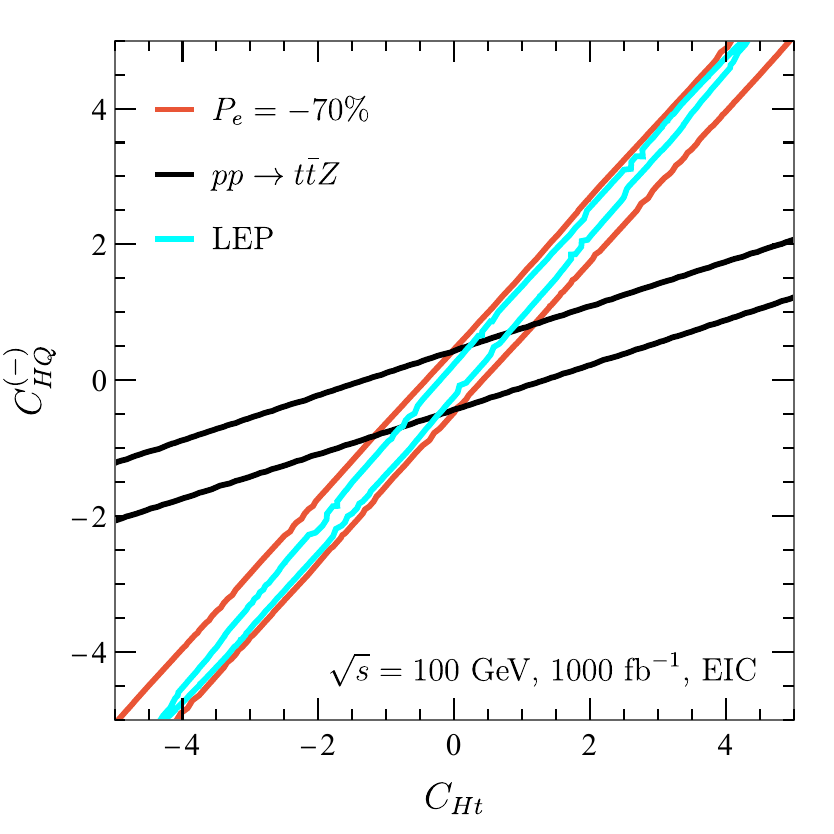}
    \includegraphics[width=0.3\linewidth]{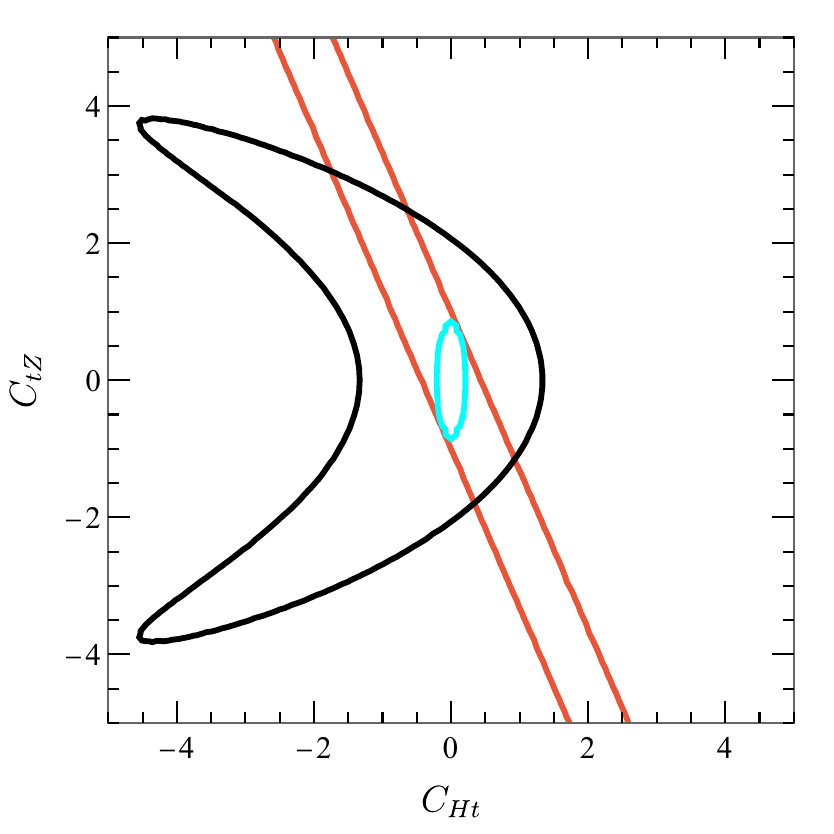}
    \includegraphics[width=0.3\linewidth]{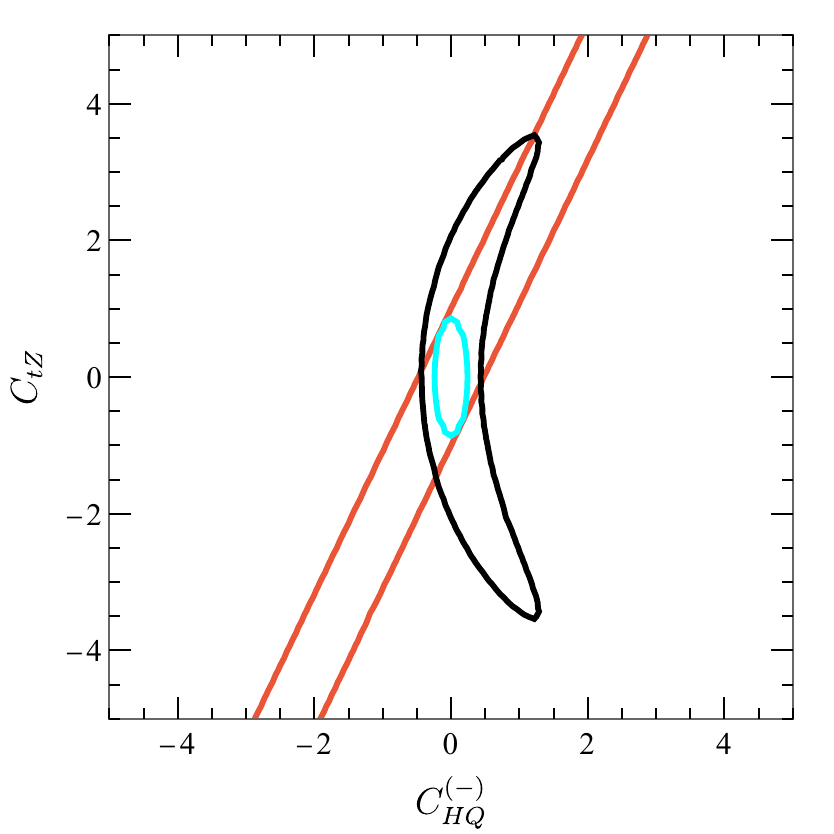}
    \caption{Similar to FIG.~\ref{fig: lhc_eic100}, but for the EIC high-luminosity phase, with an integrated luminosity of $1000~\rm fb^{-1}$.}
    \label{fig: lhc_eic100_hl}
\end{figure*}

Following recent studies~\cite{Boughezal:2022pmb}, we also include the polarized asymmetry $A_{\rm FB}$ as a complementary observable:
\begin{equation}
    A_{\rm FB} = \frac{\sigma_+ - \sigma_-}{\sigma_+ + \sigma_-}~, \label{eq:afb}
\end{equation}
where $\sigma_{\pm}$ denotes cross sections for $P_e=\pm 70\%$. The statistical uncertainty of $A_{\rm FB}$ is propagated from $\sigma_{\pm}$, leading to a corresponding $\chi^2$ term:
\begin{equation}
   \chi^2(A_{\rm FB})= \frac{(A_{\rm FB}-A_{\rm FB}^{(\rm SM)})^2}{\delta_A^2}~,
\end{equation}
where $A_{\rm FB}^{(\rm SM)}$ is the SM prediction of the asymmetry, and $\delta_A$ is the corresponding statistical uncertainty. For a consistent comparison with limits derived from cross sections, we evaluate the polarized asymmetry, $A_{\rm FB}$, at the EIC in both baseline and HL scenarios, assuming integrated luminosities of $50~{\rm fb}^{-1}$ (baseline) and $500~{\rm fb}^{-1}$ (HL). To simplify the analysis, we consider pair of operators simultaneously and apply a $\chi^2 = 2.28$ to establish 68\% confidence-level bounds. The resulting constraints for the EIC baseline and HL-phase are presented in Fig.~\ref{fig: lhc_eic100} and Fig.~\ref{fig: lhc_eic100_hl}, respectively.

As expected, the linear dependence of the cross sections on the top quark couplings produces straight-band confidence regions. The EIC shows limited sensitivity in a specific direction of the WC combinations but imposes stringent constraints along a nearly orthogonal direction. For example, in the $C_{Ht} - C_{HQ}^{(-)}$ plane, the axial-vector interaction yields tight bounds, while the vector interaction remains largely unconstrained. This behavior stems from the low momentum-transfer regime, where $Q \ll 2m_t$, causes the contribution from the vector-type operator, specifically the linear combination $\mathcal{O}_{Ht} + \mathcal{O}_{HQ}^{(-)}$, to be significantly suppressed compared to the axial-vector operator, $\mathcal{O}_{Ht} - \mathcal{O}_{HQ}^{(-)}$ (see Table~\ref{tab: xsec1}). This observation is consistent with findings at LEP, as reported in Ref.~\cite{Liu:2022vgo}. The expected limits for an electron polarization of $P_e = +70\%$ lie outside the contour ranges and are not shown, as they are significantly weaker than those for $P_e = -70\%$ (red lines) and the unpolarized (blue lines) case due to suppressed cross sections. The projected constraints from the asymmetry, $A_{\rm FB}$ (orange lines), are comparable to those in the unpolarized electron beam scenario. However, except in the $C_{Ht} - C_{HQ}^{(-)}$ plane, the orientations of the $A_{\rm FB}$ projections differ from and complement those of the unpolarized cross section. Nonetheless, the $A_{\rm FB}$ contour is significantly broader than that obtained from the cross sections for $P_e=-70\%$, indicating relatively weaker complementarity between the cross section and asymmetry measurements. The constraints are expected to improve further at the HL-phase of the EIC, as shown in Fig.~\ref{fig: lhc_eic100_hl}.

To highlight the complementary role of the EIC in probing top quark gauge couplings, we recast limits from direct measurements of inclusive and differential cross sections for $pp \to t \bar{t} Z$ production at the LHC using MadGraph5\_aMC@NLO at $\sqrt{s} = \SI{13}{TeV}$, comparing our predictions with unfolded distributions from the ATLAS~\cite{ATLAS:2023eld} and CMS~\cite{CMS:2019too,CMS:2021aly} collaborations. For operators $\mathcal{O}_{Ht}$ and $\mathcal{O}_{HQ}^{(-)}$, only linear interference with the SM is considered, as established in Ref.~\cite{BessidskaiaBylund:2016jvp}, while the dipole operator $\mathcal{O}_{tZ}$ shows negligible linear contributions due to suppression of the $\sigma^{\mu\nu} q_\nu$ interaction by the $Z$ boson momentum and an accidental cancellation between $gg$ and $\bar{q}q$ channels, with quadratic effects dominating, consistent with ATLAS results~\cite{ATLAS:2023eld}. Consequently, both real and imaginary components of the dipole operators contribute to $t\bar{t}Z$ production, though we exclude imaginary contributions for comparison with our results by requiring all WCs to be real. We constrain $ t \bar{t}Z$ couplings by analyzing inclusive and differential $Z$ boson transverse momentum distributions, finding the ATLAS results~\cite{ATLAS:2023eld} provide the strongest LHC constraints, depicted as black curves in Figs.~\ref{fig: lhc_eic100} and~\ref{fig: lhc_eic100_hl}.  Electroweak precision measurements from the LEP, shown as cyan lines in the same figures, impose the tightest constraints indirectly~\cite{Liu:2022vgo}, yet the precision in the EIC's HL-phase, as shown in Fig.~\ref{fig: lhc_eic100_hl}, is anticipated to be comparable, highlighting the critical role of the EIC’s DIS process in complementing these measurements.

Before closing this section, we note that our results exhibit a strong dependence on the EFT renormalization scale $\mu_{\text{EFT}}$, as the WCs of irrelevant operators are assumed to vanish at $\mu_{\text{EFT}}$. However, RGE activates these operators, causing each coefficient $r_i$ to influence other $r$ values, leading to unstable cross-section corrections unless all operator contributions are included, as discussed in Ref.~\cite{Maltoni:2016yxb}. A more robust approach, as suggested in Ref.~\cite{Maltoni:2016yxb}, involves presenting two-dimensional contours after marginalizing over all irrelevant operators. In this initial exploration of constraining top quark couplings at the EIC, we focus exclusively on the most relevant operators, leaving marginalization over irrelevant operators beyond the scope of this study.

\emph{Conclusions.---}
In this Letter, we propose measuring top quark electroweak couplings indirectly through the DIS process at the EIC within the SMEFT framework, incorporating next-to-leading order electroweak corrections. Our analysis shows that the DIS cross section is significantly enhanced for a left-handed electron beam compared to right-handed polarized scenario. By comparing these findings with direct measurements of $pp\to t\bar{t}Z$ at the LHC, we demonstrate that DIS measurements at the EIC can play a pivotal role in complementing constraints on top quark couplings, effectively resolving parameter space degeneracies. While the baseline integrated luminosity at the EIC and LHC yield constraints weaker than electroweak precision measurements from LEP, the high-luminosity phase of the EIC is expected to achieve limits comparable to LEP, underscoring the EIC's critical role in advancing our understanding of top quark physics.

\emph{Note Added:}
During the finalization of this manuscript, Ref.~\cite{Bellafronte:2025ubi} was published as a preprint, which examines electroweak corrections to the DIS process arising from top-electron four-fermion interactions.

\vspace{3mm}
\begin{acknowledgments}

We would like to deeply memorize our friend Cen Zhang, whose UFO file enabled NLO precision in our EIC simulations. We also thank Jiayin Gu, Haolin Li, and Kun-Feng Lyu for insightful discussions on EFT renormalization. X.J. was supported in part by the National Natural Science Foundation of China under grant No.~12342502. Y.L. was supported by  the National Key R$\&$D Program of China (grant 2023YFE0117200) and the National Natural Science Foundation of China (Nos. 12105013). B.Y. was supported in part by the National Science Foundation of China under Grant No.~12422506 and CAS under Grant No.~E429A6M1. The authors gratefully acknowledge the valuable discussions and insights provided by the members of the Collaboration on Precision Tests and New Physics (CPTNP).
\end{acknowledgments}

\bibliography{DISEIC}

\begin{widetext}
\appendix
\section{The results for higher central energy} \label{sec:appendix1}

In this appendix, we investigate an alternative EIC configuration with $E_e = 18~\rm GeV$ and $E_p = 275~\rm GeV$, achieving a center-of-mass energy of approximately 140 GeV. We consider three electron polarization scenarios. The linear corrections to the SM cross sections, based on our calculations, are summarized in Table~\ref{tab: xsec2}.

\begin{table*}[htbp]
\centering
\begin{tabular}{cccccc}
\hline
 $P_e$ & $\sigma_{\rm SM}$[pb] & $r_{Ht}$[pb] & $r_{HQ}^{(-)}$[pb] & $r_{tZ}$[pb] \\
\hline
$-70\%$ & $208$ & $7.43\times 10^{-2}$ & $-6.71\times 10^{-2}$ & $3.26\times 10^{-2}$ \\
\hline
$0$ & $200$ & $3.51\times 10^{-2}$ & $-3.16\times 10^{-2}$ & $1.40\times 10^{-2}$  \\
\hline
$+70\%$ & $193$ & $-3.86\times 10^{-3}$ & $3.57\times 10^{-3}$ & $-4.09\times 10^{-3}$ \\
\hline
\end{tabular}
\caption{Similar to Table~\ref{tab: xsec1}, but with $\sqrt s = 140~\rm GeV$.}\label{tab: xsec2}
\end{table*}

\begin{figure*}[h]
    \centering
    \includegraphics[width=0.3\linewidth]{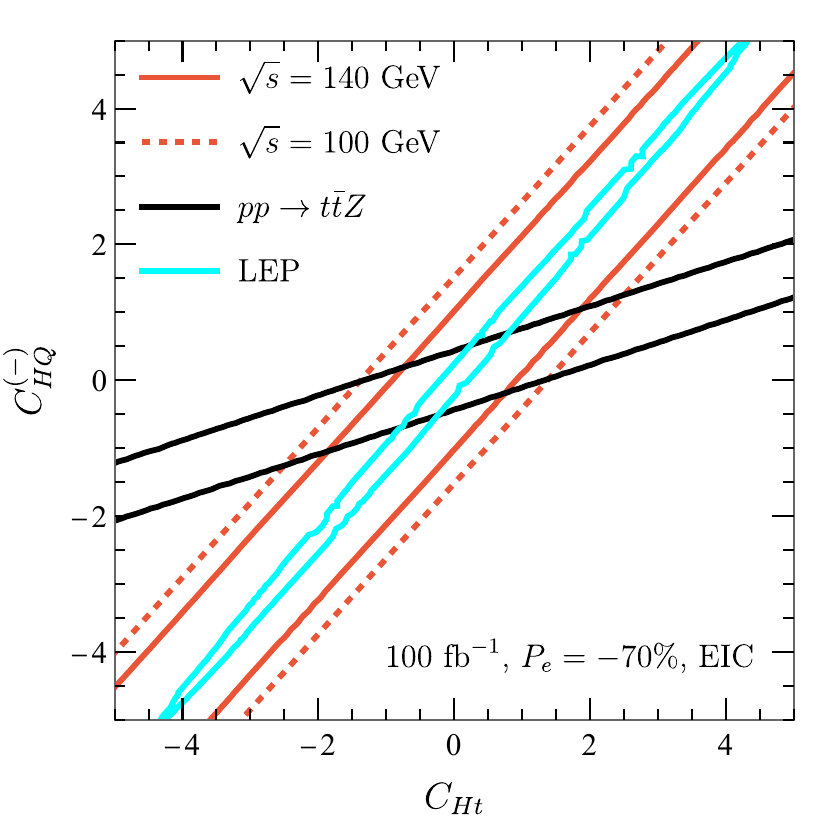}
    \includegraphics[width=0.3\linewidth]{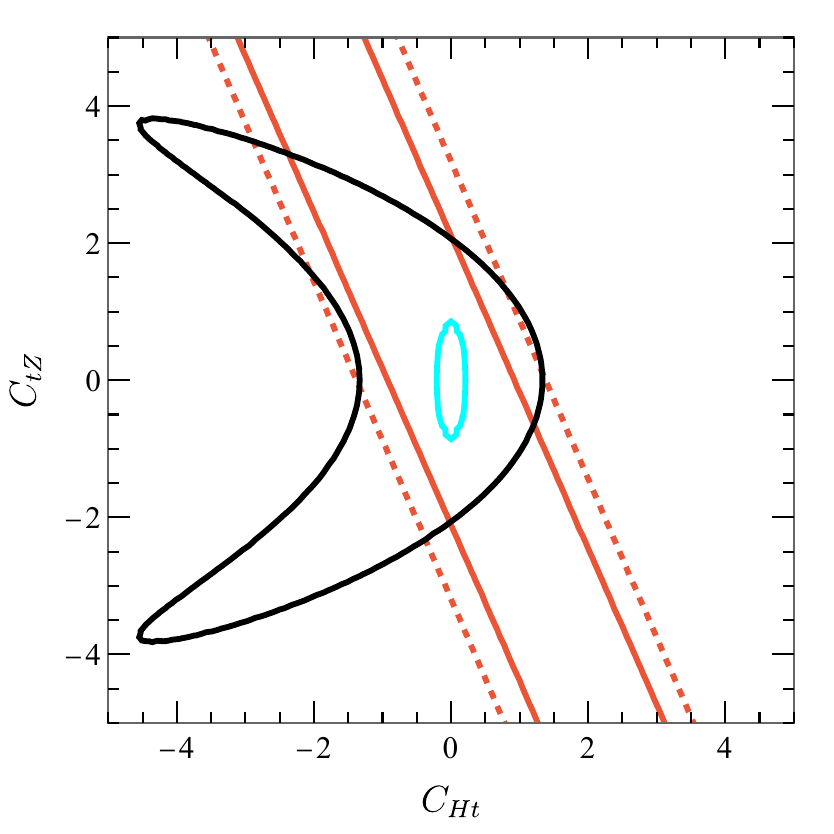}
    \includegraphics[width=0.3\linewidth]{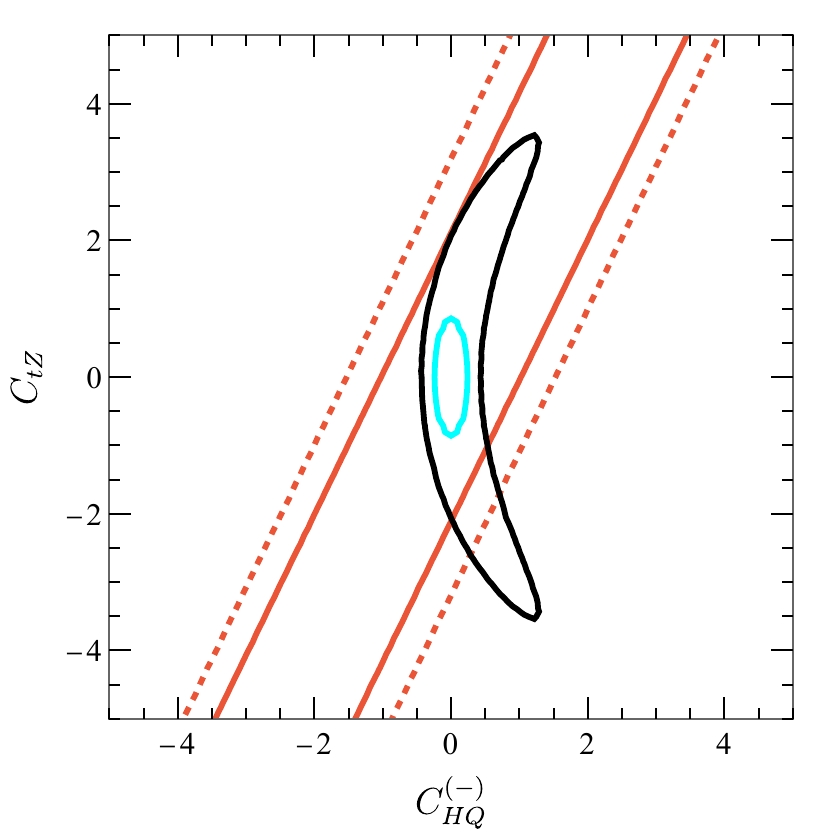}
    \caption{Expected constraints on top quark electroweak couplings at 68\% confidence level from the EIC at $\sqrt{s} = 140~\rm GeV$ (solid lines) and $\sqrt{s} = 100~\rm GeV$ (dashed lines), with an integrated luminosity of $100~\rm fb^{-1}$.}
    \label{fig: lhc_eic140}
\end{figure*}

Figure~\ref{fig: lhc_eic140} displays the projected constraints on top quark electroweak couplings at a 68\% confidence level from the EIC at $\sqrt{s} = 140~\rm GeV$ (solid lines) and $\sqrt{s} = 100~\rm GeV$ (dashed lines), with an integrated luminosity of $100~\rm fb^{-1}$. As anticipated, the higher center-of-mass energy results in more stringent constraints on top quark couplings.
\end{widetext}
 
\end{document}